\documentclass[a4paper,11pt]{article}%
\usepackage{amssymb}
\usepackage{amsfonts}
\usepackage{amsmath,bbm}
\usepackage[nohead]{geometry}
\usepackage{graphicx}
\RequirePackage[colorlinks,citecolor=blue]{hyperref}
\usepackage{algorithm}
\usepackage{algorithmic}
\usepackage{caption}
\usepackage{subcaption}
\usepackage{natbib}

\newcommand{\Rmnum}[1]{\expandafter\romannumeral #1}

 
\newcommand{\RNum}[1]{\uppercase\expandafter{\romannumeral #1\relax}}

\def\m{\mathcal}
\def\mb{\mathbb}
\def\mc{\mathcal}
\def\mb{\mathbb}
\def\mr{\mathrm}

\newcommand{\ind}{\mathbbm{1}}

\def\T{{ \mathrm{\scriptscriptstyle T} }}

\title{Efficient Bayesian shape-restricted function estimation with constrained Gaussian process priors }

\author{Pallavi Ray \thanks{pallaviray@stat.tamu.edu} \quad \quad Debdeep Pati \thanks{debdeep@stat.tamu.edu} \quad \quad Anirban Bhattacharya \thanks{anirbanb@stat.tamu.edu}}

\date{\it Department of Statistics, Texas A\&M University \\ \it Address: 3143 TAMU, College Station, TX 77843, USA}

\begin{document}
\maketitle

\abstract{
 
This article revisits the problem of Bayesian shape-restricted inference in the light of a recently developed approximate Gaussian process that admits an equivalent formulation of the shape constraints in terms of the basis coefficients.  We propose a strategy to efficiently sample from the resulting constrained posterior by absorbing a {\em smooth relaxation} of the constraint in the likelihood and using circulant embedding techniques to sample from the unconstrained {\em modified prior}. We additionally pay careful attention to mitigate the computational complexity arising from updating hyperparameters within the covariance kernel of the Gaussian process. The developed algorithm is shown to be accurate and highly efficient in simulated and real data examples. \\
}


\section{Introduction}\label{sec:intro}

In diverse application areas, it is often of interest to estimate a function nonparametrically subject only to certain constraints on its shape. 
Typical examples include (but are not limited to) monotone dose-response curves in medicine \citep{kelly1990monotone}, concave utility functions in econometrics \citep{meyer1968consistent}, increasing growth curves, non-increasing survival function or `U'-shaped hazard function \citep{reboul2005estimation} in survival analysis, computed tomography \citep{prince1990constrained}, target reconstruction \citep{lele1992convex}, image analysis \citep{goldenshluger2006recovering}, queueing theory \citep{chen1993dynamic}, and circuit design \citep{nicosia2008evolutionary}.


A Bayesian framework offers a unified probabilistic way of incorporating various shape constraints and accordingly there is a large literature devoted to Bayesian shape constrained estimation. A general approach is to expand the unknown function in a basis and translating the functional constraints to linear constraints in the coefficient space. Some representative examples include piecewise linear models \citep{neelon2004bayesian,cai2007bayesian}, Bernstein polynomials \citep{curtis2009variable}, regression splines \citep{meyer2011bayesian}, penalized spines \citep{brezger2008monotonic}, 
cumulative distribution functions \citep{bornkamp2009bayesian}, and restricted splines \citep{shively2011nonparametric} used as the basis. \cite{maatouk2017gaussian} recently exploited a novel basis representation to equivalently represent various shape restrictions such as boundedness, monotonicity, convexity etc as non-negativity constraints on the basis coefficients. Although originally developed in the context of computer model emulation, the approach of \cite{maatouk2017gaussian} is broadly applicable to general shape constrained problems. \cite{zhou2018revisiting} adapted their approach to handle a combination of shape constraints in a nuclear physics application to model the electric form factor of a proton. The main idea of \cite{maatouk2017gaussian} is to expand a Gaussian process using a first or second order exact Taylor expansion, with the remainder term approximated using linear combinations of compactly supported triangular basis functions. A key observation is that the resulting approximation has the unique advantage of enforcing linear inequality constraints on the function space through an {\em equivalent} linear constraint on the basis coefficients. In terms of model fitting under a standard Gibbs sampling framework, this necessitates sampling from a high-dimensional truncated multivariate normal (tMVN) distribution.

The problem of sampling from a tMVN distribution is notoriously challenging in high dimensions and a number of solutions have been proposed in the literature. Existing Gibbs samplers for a tMVN distribution sample the coordinates one-at-a-time from their respective full conditional truncated univariate normal distributions \citep{geweke1991efficient, kotecha1999gibbs, damien2001sampling, rodriguez2004efficient}.  While the Gibbs sampling procedure is entirely automated, such one-at-a-time updates can lead to slow mixing, especially if the variables are highly correlated. More recently, \cite{pakman2014exact} proposed a Hamiltonian Monte Carlo (HMC) algorithm which has drastically improved the speed and efficiency of sampling from tMVNs. However, implementing this algorithm within a larger Gibbs sampler can still lead to inefficiencies if the sample size is large. The second and third authors of this article encountered this challenge in \cite{zhou2018revisiting} with a sample size greater than 1000. A related issue which contributes to the complexity is the $\mc O (N^3)$ computation and $\mc O (N^2)$ storage requirements for inverting and storing a general $N \times N$ covariance matrix. 


In this article, we propose a novel algorithm to exploit additional structure present in the tMVN distributions arising in the aforesaid shape-constrained problems using the basis of \cite{maatouk2017gaussian}. Our approach is based on a novel combination of elliptical slice sampling (ESS; \cite{murray2010elliptical}), circulant embedding techniques, and smooth relaxations of hard constraints. We additionally use Durbin's recursion to efficiently update hyperparameters within the covariance kernel of the parent Gaussian process. We analyze the per-iteration complexity of the proposed algorithm with and without hyperparameter updates and illustrate through simulated and real data examples that the proposed algorithm provides significant computational advantages while retaining the statistical accuracy. \texttt{R} code to implement the proposed algorithm for monotone and convex function estimation is provided at \href{URL}{https://github.com/raypallavi/BNP-Computations}. We note that our algorithm and code be trivially adapted to other basis functions.

The rest of the paper is organized as follows.  In \S \ref{sec:methods}, we revisit the Bayesian shape constrained function estimation problem and describe the novel algorithm for inference. The algorithm is specialized to estimating monotone and convex functions \S \ref{sec:app} with illustrations in \S \ref{sec:sims}. We conclude with a real data application in \S \ref{sec:real}. Various algorithmic and implementation details are provided in an Appendix.

\section{Algorithm development}\label{sec:methods}

Consider the problem of sampling from a distribution having the following form:
\begin{align}\label{full-cond-xi}
p(\xi) \propto \,\, \exp \bigg\{ - \frac{1}{2 \sigma^2} \|Z - \Phi \xi \|^2 \bigg \} \, \exp \bigg\{ - \frac{1}{2 \tau^2} \xi^\T K^{-1} \xi \bigg \} \,\, \ind_{\mc C_{\xi}}(\xi), \quad \xi \in \mathbb{R}^N,
\end{align}
where, $Z \in \mb R^n$, $\Phi \in \mb R^{n \times N}$ with $N \le n$, $\mc C_{\xi} \subset \mathbb{R}^N$ is determined by a set of linear inequality constraints on $\xi$, and $K$ is positive definite matrix. While our methodology generally applies to any such $K$, we are specifically interested in situations where $K$ arises from the evaluation of a stationary covariance kernel on a regular grid. 

The distribution \eqref{full-cond-xi} arises as a conditional posterior of basis coefficients in many Bayesian nonparametric regression problems where linear shape constraints (such as monotonicity, convexity, or a combination of these; \cite{zhou2018revisiting}) on the regression function are present, and a constrained Gaussian process prior is placed on the coefficient vector $\xi$ in an appropriate basis representation. Sampling from the density \eqref{full-cond-xi} is then necessitated within a larger Gibbs sampler to fit the said constrained regression model. 

Specifically, suppose we observe response-covariate pairs $(y_i, x_i) \in \mb R \otimes \mb R^d$ for $i = 1, \ldots, n$, related by the Gaussian regression model
\begin{eqnarray}\label{eq:model}
y_i = f(x_i) + \epsilon_i,  \quad  \epsilon_i  \sim \mbox{N}(0, \sigma^2), \quad i = 1,\dots, n
\end{eqnarray}
where the unknown regression function $f$ is constrained to lie in some space $\m C_f$, a subset of the space of all continuous functions on $[0, 1]^d$. When $\m C_f$ corresponds to the space of monotone or convex functions, \cite{maatouk2017gaussian} identified a novel basis representation for $f$ which allowed {\em equivalent representations} of the aforesaid constraints in terms of the basis coefficients $\xi$ restricted to the positive orthant: 
\begin{align}\label{eq:c_xi}
\mc C_{\xi} :\,= \mc C_{\xi}^N = \bigg\{\xi\in \mathbb{R}^{N}: ~ \xi_j \ge 0, \ j = 1, \dots, N \bigg\}. 
\end{align}
We provide more details on their basis representation in \S \ref{mon} and \S \ref{con}. 
They also considered the case where the regression function is globally bounded between two constants. See also \cite{zhou2018revisiting} where a combination of interpolation, monotonicity, and convexity constraints can be equivalently expressed in terms of linear constraints on the coefficients. 


Relating the basis coefficients $\xi$ with the function values and (or) its derivatives, \cite{maatouk2017gaussian} proposed a constrained Gaussian prior on $\xi$. If the function $f$ was unconstrained, then a Gaussian process (GP) prior on $f$ induces a Gaussian prior on $\xi$, aided by the fact that derivatives of GP are again GPs provided the covariance kernel is sufficiently smooth. A natural idea then is to restrict the induced prior on $\xi$ to the constrained region $\m C_\xi$, 
$$
\pi(\xi) \,\propto\, \m N(\xi; 0, \tau^2 K) \, \ind_{\m C_\xi}(\xi), 
$$
which is precisely the specification of \cite{maatouk2017gaussian}. The density in equation \eqref{full-cond-xi} is then recognized as the conditional posterior of $\xi$. 

In what follows, we shall additionally assume that $K_{jj'} = k(u_j - u_{j'})$ for a positive definite function $k$ and a set of uniform grid points $\{u_j\}$. For example, if $d = 1$, we have $u_j = j/N$ for $j = 0,1, \ldots, N$. This is a slight departure from \cite{maatouk2017gaussian} in the monotone and convex case. Since the derivatives of a stationary GP is generally non-stationary, so is their induced prior on $\xi$ from a parent stationary GP on $f$. We instead directly place a stationary GP on an appropriate derivative of $f$, which results in $K$ having a form as above. While there is little difference between the two approaches operationally, there is a large computational benefit for our approach, as we shall see below. 

Returning to \eqref{full-cond-xi}, a simple calculation yields that $p(\xi)$ is a truncated normal distribution, specifically, 
$$
\m N_N \Big((\Phi^\T \Phi / \sigma^2 + K^{-1} / \tau^2)^{-1} \Phi^\T Y, (\Phi^\T \Phi / \sigma^2 + K^{-1} / \tau^2)^{-1} \Big),
$$
truncated to $\m C_{\xi}$. While one can use off-the-shelf samplers for tMVNs  \citep{pakman2014exact} to sample from the above, the intrinsic complexity of sampling from tMVNs coupled with the computation and storage of the inverse of the kernel matrix $K$ contributes to the challenges of executing this sampling step for large $N$. In particular, $(\Phi^\T \Phi / \sigma^2 + K^{-1} / \tau^2)$ keeps changing over each MCMC iteration with new updates of $\sigma$ and $\tau$, which requires an $N \times N$ matrix inversion at each iteration while applying any of the existing algorithms. The usual Sherman--Morrison--Woodbury matrix inversion trick does not render beneficial in this case. Barring issues with matrix inversions, implementation of such algorithm will be expensive in terms of storage. In addition, if there are unknown hyperparameters in the covariance kernel that get updated at each iteration within a larger MCMC algorithm, either the inversion has to take place at each step, or one has to pre-store a collection of $K^{-1}$ on a fine grid for the hyperparameters.

In this article, we present a different approach to sample from the density $p$ in \eqref{full-cond-xi} which entirely avoids matrix inversions. Our approach is based on three basic building blocks: (i) approximating the indicator function in $p$ with a smooth approximant, (ii) a novel use of elliptical slice sampling \citep{murray2010elliptical} to avoid sampling from truncated non-Gaussian distribution, and (iii) using highly efficient samplers based on the fast Fourier transform for stationary GPs on a regular grid \citep{wood1994simulation}. We describe the details below, starting with a brief review of elliptical slice sampling. 

The elliptical slice sampler is a general technique for sampling from posterior distributions of the form,
$$
p(\xi) \,\propto\, L(\xi) \, \mc N (\xi; 0, \Sigma)
$$
proportional to the product of a zero-mean multivariate Gaussian prior with a general likelihood function $L(\cdot)$. In this context, Metropolis--Hastings proposals 
$$\xi' =  \rho \, \nu_e + \sqrt{1-\rho^2} \,\xi, \quad \nu_e \sim \m N(0, \Sigma)$$
for $\rho \in [-1, 1]$ are known to possess good empirical \citep{bernardo1998regression} and theoretical \citep{cotter2013mcmc} properties. 
The elliptical slice sampler presents an adaptive and automated way to tune the step-size parameter which guarantees acceptance at each step. Specifically, a new location on the randomly generated ellipse determined by the current state $\xi$ and the auxiliary draw $\nu_e$ is produced according to 
\begin{align}\label{ess}
\xi' = \nu_e \sin \theta + \xi \cos \theta 
\end{align}
where the angle $\theta$ is uniformly generated from a $[ \theta_{\min}, \theta_{\max}]$ interval which is shrunk exponentially fast until an acceptable state is reached. The only requirement is the ability to evaluate $L$ at arbitrary points, rendering the approach broadly applicable. 

Turning to \eqref{full-cond-xi}, note however that the elliptical slice sampler isn't immediately applicable as we have a truncated normal prior. As a simple fix-up, we approximate the indicator function  $\mathbbm{1}_{\mc C_{\xi}} (\cdot)$ in \eqref{full-cond-xi} by a suitable smooth function. Specifically, assuming $\mc C_{\xi}$ has the same structure as in \eqref{eq:c_xi}, we use sigmoid-like approximations $ \ind_{(0, \infty)}(x) \approx (1 + e^{- \eta x})^{-1}$ for large $\eta > 0$ to obtain a smooth approximation $\mb J_{\eta}(\cdot)$ to $\mathbbm{1}_{\mc C_{\xi}} (\cdot)$ as
$$ 
\mathbbm{1}_{\mc C_{\xi}} (\xi) \approx \mb J_{\eta}(\xi) = \prod_{j=1}^{N} \frac{e^{\eta \xi_j}}{1 + e^{\eta \xi_j}}.
$$
The parameter $\eta$ controls the quality of the approximation; higher the value of $\eta$, better is the approximation. Experimenting across a large number of simulation scenarios, we find that $\eta = 50$ already provides a highly accurate approximation for all practical purpose. 

With $\mb J_{\eta}(\xi)$ defined like this, let us define
\begin{eqnarray} \label{cond_post_xi}
\widetilde{p}(\xi \mid -) & \propto & \exp \bigg\{ - \frac{1}{2 \sigma^2} \|Y - \Phi \xi \|^2 \bigg \} \, \exp \bigg\{ - \frac{1}{2 \tau^2} \xi^\T K^{-1} \xi \bigg \} \, \mb J_{\eta}(\xi) \nonumber \\
& = & \bigg[ \exp \bigg\{ - \frac{1}{2 \sigma^2} \|Y - \Phi \xi \|^2 \bigg \} \, \mb J_{\eta}(\xi) \bigg] \, \exp \bigg\{ - \frac{1}{2 \tau^2} \xi^\T K^{-1} \xi \bigg \} \nonumber \\
& = & \bigg[ \exp \bigg\{ - \frac{1}{2 \sigma^2} \|Y - \Phi \xi \|^2 \bigg \} \, \bigg\{ \prod_{j=1}^{N} \frac{e^{\eta \xi_j}}{1 + e^{\eta \xi_j}} \bigg\} \bigg] \, \exp \bigg\{ - \frac{1}{2 \tau^2} \xi^\T K^{-1} \xi \bigg \}.
\end{eqnarray}
Our goal now is to sample from $\widetilde{p}$ which we shall consider as approximate samples from $p$. 

Now we can apply ESS to draw samples from $\widetilde{\pi}(\xi \mid -)$, since treating the quantity in the square brackets in (\ref{cond_post_xi}) as ``redefined likelihood", $\xi$ has an (untruncated) multivariate Gaussian prior, which we call as the ``working prior". So one just needs to draw from the ``working prior" distribution and compute the logarithm of the ``redefined likelihood" function. In our case, computing the log-likelihood function has computational cost of $\mc O (nN)$ and we are to sample $\nu_e \sim \mc N(0, \tau^2 \, K)$, which is usually $\mc O(N^3)$. Note that these computational complexities correspond to a single iteration of the MCMC sampler.

Now under the assumption that the covariance matrix $K$ is obtained from a regular grid in $[0, 1]$, sampling from the ``working prior" is same as simulating realizations of a stationary Gaussian Process on a regular grid in $[0, 1]$. Such a covariance matrix is known to have a Toeplitz structure and the simulation can be carried out using the sampling scheme developed by \cite{wood1994simulation} which reduces the overall complexity of the algorithm to a commendable extend. The details of this algorithm is discussed in the following section.

\subsection{Sampling from the prior distribution of $\xi$}

Sampling from the ``working prior" distribution requires sampling from a stationary GP on a regular grid in $[0,1]$ with a Toeplitz structure of the covariance matrix. In such settings, the algorithm of \cite{wood1994simulation} based on clever embedding techniques can be readily applied. In particular, they exploit the discrete fast Fourier transform twice to offer substantially reduced compared cost. We briefly discuss some of the key ingredients of the algorithm.

The goal is to sample a random vector of the form 
$$Z = \bigg( Z \big(0 \big), Z \bigg(\frac{1}{m} \bigg), Z \bigg(\frac{2}{m} \bigg), \cdots, Z \bigg(\frac{m-1}{m} \bigg) \bigg)^\T $$
from a mean-zero Gaussian random process on each of the grid points $$\bigg \{ 0, \frac{1}{m}, \frac{2}{m}, \cdots, \frac{m-1}{m} \bigg \} \hspace{2mm} \equiv \hspace{2mm} \bigg \{ u_j : u_j = \frac{j}{m} ; 0 \leq j < 1 \bigg \}, \hspace{3mm} m \geq 1 $$ with covariance function $\gamma : \mb R \to \mb R$. Then $Z \sim \mc N_m(0, G)$, where

\[ G = 
\begin{bmatrix}
\gamma(0)&\gamma \big(\frac{1}{m} \big)&\cdots&\gamma \big(\frac{m-1}{m} \big) \\
\gamma \big(\frac{1}{m} \big)&\gamma(0)&\cdots &\gamma \big(\frac{m-1}{m} \big) \\
\vdots & \vdots & \ddots & \vdots\\
\gamma \big(\frac{m-1}{m} \big)&\gamma \big(\frac{m-2}{m} \big)&\cdots &\gamma(0)
\end{bmatrix}
\]
It is to be noted that $G$ is a Toeplitz matrix, which is equivalent to $\tau^2 K$ in our notation.

There are two basic steps of this method:
\begin{itemize}
	\item[1] Embedding $G$ in a circulant covariance matrix $C$ of order $d \times d, \text{where} \hspace{3mm} d = 2^g$, for some integer $g$ and $d \geq 2(m-1)$. The circulant matrix is formed such a way that by construction, $C$ is symmetric and the $m \times m$ submatrix in the top left corner of $C$ is equal to $G$. For details on the embedding technique, one can refer to \cite{wood1994simulation}. 
	\item[2] Using fast Fourier transform \textbf{twice} to generate $Z = \big( Z_0, Z_1, \cdots, Z_{d-1} \big) \sim \mc N_d(0, C)$. Then due to appropriate construction of $C$, $\big( Z_0, Z_1, \cdots, Z_{m-1} \big) \sim \mc N_m(0, G)$.
\end{itemize} 
Thus the problem of sampling from $\mc N_m(0, G)$ changes to sampling from $\mc N_d(0, C)$, which is much more efficient due to double implementation of super efficient FFT algorithm in the procedure. This is done with computational complexity of $\mc O (d \log d)$.

It is to be noted that,  the circulant matrix $C$ is not guaranteed to be positive definite.  In practice, it is usually possible to choose $d$ large enough such that $C$ is nonnegative definite \citep{wood1994simulation} and in such cases the procedure is {\em exact} in principle. But if such a $d$ cannot be found or if it is too large to be practicable, then an {\em approximate} method is used to make $C$ nonnegative definite, where the matrix $C$ is split into nonnegative definite and nonpositive definite parts; 
For more details on the approximate method one can consult \cite{wood1994simulation}. For our algorithm, we prefer the {\em exact} method and use an appropriate $d$ that satisfies all the necessary conditions. Applying this algorithm a sample is drawn from $\mc N_N (0, \tau^2 \, K)$ with $\mc O (N \log N)$ (for each MCMC iteration) computations, instead of $\mc O (N^3)$ computations, and is used as the proposal $\nu_e$ in ESS given by (\ref{ess}).

\subsection{Algorithm}

We implemented our algorithm with $K$ as a stationary M{\'a}tern kernel with smoothness parameter $\nu > 0$ and length-scale parameter $\ell > 0$. Our method takes design points $X$, observations $Y$, M{\'a}tern kernel parameters $\nu$ and $\ell$, $\eta$ as in \eqref{cond_post_xi}, dimension of the random coefficients $N$ and number of posterior samples $n_0$ as inputs and gives $n_0$ many posterior samples of $\xi$ as output.

\begin{algorithm}
	\caption{Efficient algorithm to draw posterior samples of $\xi$}
	\label{algo0}
	\begin{algorithmic}
		\REQUIRE $X$, $Y$, $\nu$, $\ell$, $\eta$, $N$, $\tau^2$, $\sigma^2$ and $n_0$
		\STATE Using $N$, calculate $u_j = j/N \, , \,\, j=0, \ldots,N $; 
		\STATE Using $X$ and $u_j$'s form basis matrix $\Phi$
		\STATE Using $\nu$, $\ell$ and $u_j$'s form covariance matrix $K$ 
	    \STATE Initialize : $\xi^{(0)}$
		\FOR{$t=1$ \TO $n_0$}
		\STATE Sample $\nu_e \sim \mc N(0, \tau^2 \, K)$ using simulation scheme by \cite{wood1994simulation}.
		\STATE Sample $\xi^{(t)}$ using $\nu_e$, $\eta$ and $\sigma^2$ following ESS scheme by \cite{murray2010elliptical}.
		\ENDFOR
		
		\ENSURE Posterior samples of $\xi$ of size $n_0$.
	\end{algorithmic}
\end{algorithm}

The computation cost for drawing a random sample from the prior distribution usually dominates. But that is not the case here. Since $n > N$, computational cost for computing the log-likelihood, using ESS scheme, dominates which leads to computational complexity of $\mc O \big(n N \big)$, for each MCMC iteration. 

\subsection{Updating hyperparameters}

As already discussed, updating the hyperparameters present in the covariance matrix $K$ is computationally challenging in the absence of any structure in $K$. Any likelihood-based method for updating $\nu$ and $\ell$ (e.g. Metropolis--Hastings)  requires computing $K^{-1}$ which leads to $\mc O(N^3)$ computational steps and $\mc O(N^2)$ storage. Hence the computational complexity per MCMC iteration of Algorithm~\ref{algo0} is always bounded above by $\mc O(N^3)$. 

However, substantial speed-up is possible in our case as $K$ is a symmetric positive-definite Toeplitz matrix. We turn to a non-trivial but effective approach of finding $K^{-1}$ utilizing inverse Cholesky factor of $K$ using Durbin's recursion algorithm \citep{golub2012matrix} which has a computational complexity of $\mc O (N^2)$. The columns of the inverse Cholesky factor of $K$ is obtained by solving Yule--Walker systems.  Durbin recursion is a procedure of recursively finding the solution to a system of equations involving a Toeplitz matrix, in particular, it is applicable to Yule--Walker systems. Given real numbers $r_0, r_1, \ldots, r_{M-1}$ with $r_0 = 1$ such that $T = \big(r_{|i-j|} \big) \in \mb R^{M \times M}$ is positive definite then Durbin's algorithm computes $u \in \mb R^M$ as a solution of the Yule--Walker problem:
$$T \, u = - (r_1, \ldots, r_{M-1})^\T$$
For more details on Durbin's recursion for solving Yule--Walker equation,  refer to \cite{golub2012matrix}.

Now suppose, we have the Cholesky factor $R$ such that $R^\T R = T$ where $R$ is an upper-triangular matrix and the inverse Cholesky factor is given by $R^{-1}$. Therefore, $T R^{-1} = R^\T$ and noting that $R^\T$ is lower-triangular, it is enough to solve only the upper-triangular part of $R^{-1}$. The first $h$ elements of the $h^{th}$ column of $R^{-1}$ can be found as a solution of 
$$T_h \, u^{(h)} = -r^{(h)} \, , \,\,\, h=1,\ldots,M$$ 
where $T_h$ is the $h \times h$ principal submatrix of $T$, $u^{(h)}$ is $h$-dimensional vector of solutions and $r^{(h)} = (r_1, \ldots, r_h)^\T$ and each of these $M$ equations can be solved using Durbin's algorithm mentioned above. Note that, the $h^{th}$ column of $R^{-1}$ denoted by $(R^{-1})_h$ is then given by:
\[ (R^{-1})_h = 
\begin{bmatrix}
E_h & O_{h \times (M-h)}\\
O_{(M-h) \times h} & O_{(M-h) \times (M-h)}
\end{bmatrix} 
\begin{bmatrix}
u^{(h)} \\
O_{(M-h) \times h}
\end{bmatrix}
\]
where $E_h$ is an \textit{exchange matrix} of order $h$ with anti--diagonal elements all ones and other elements are all zeros.
This approach requires $\mc O(M^2)$ computations to find $R^{-1}$. 

We considered continuous uniform priors on compactly supported intervals on both $\nu$ and $\ell$, independently of each other. Updating $\nu \mid \xi, -$ and $\ell \mid \xi, -$ using Metropolis--Hastings requires to compute acceptance ratio which involves computation of $\xi^\T K^{-1} \xi$ and $|K|^{-1/2}$ for proposal and current combinations of $(\nu, \ell)$. Using Durbin's algorithm, we can find $S$ such that $({S^{-1}})^\T S^{-1} = K$ and then $\xi^\T K^{-1} \xi = (S^\T \xi)^\T (S^\T \xi) = \sum_{j=1}^{N} v_j$ where $v = S^\T \xi$ and $|K|^{-1/2} = \prod_{j=1}^{N} S_{jj}$. Evidently, computation for $S$ dominates and Durbin's algorithm allows us to update the hyperparameters in $\mc O(N^2)$ computations for each iteration within an MCMC algorithm. Thus per-iteration computational complexity for Algorithm~\ref{algo0} combined with this hyperparameter update technique remains $\mc O(nN)$ as before. 

\begin{figure}[H]
	\centering
	\begin{subfigure}{.5\textwidth}
		\centering
		\includegraphics[scale=0.3]{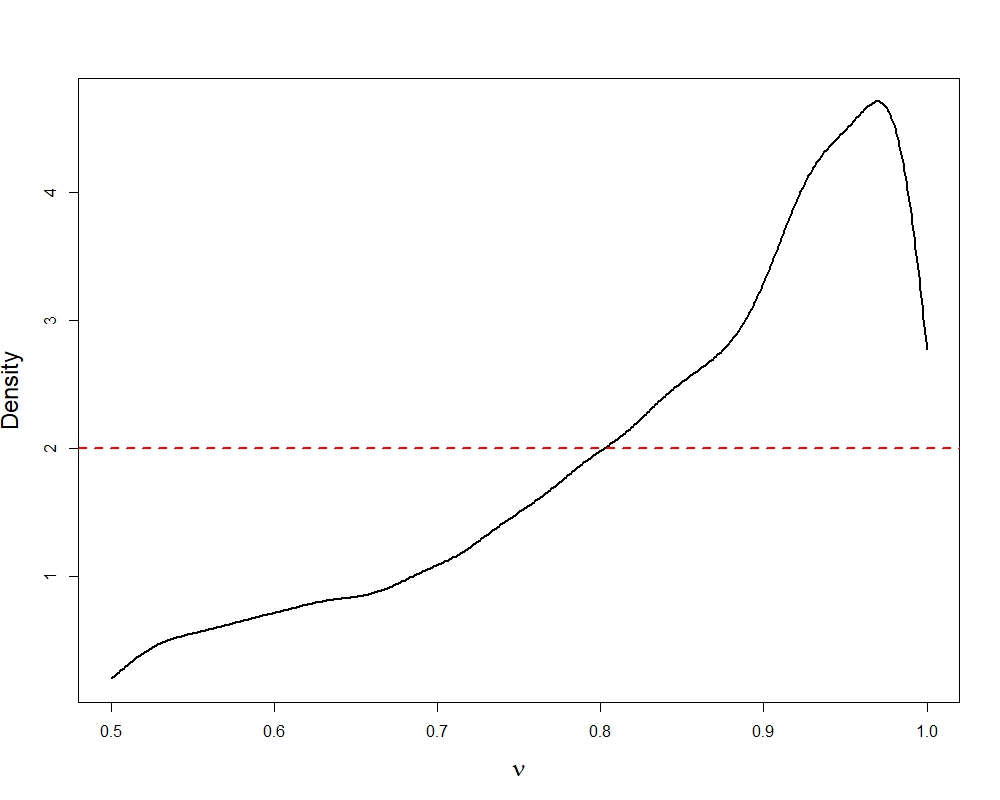}
	\end{subfigure}%
	\begin{subfigure}{.5\textwidth}
		\centering
		\includegraphics[scale=0.3]{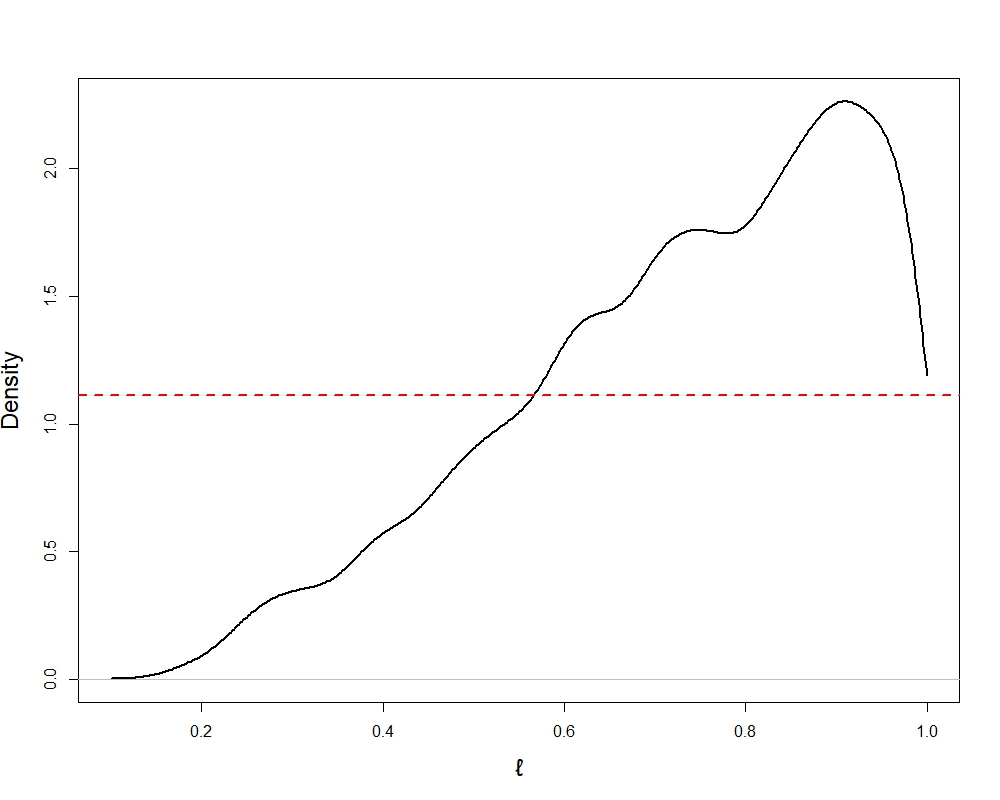}
	\end{subfigure}
	\caption{{\em Posterior density plots of the hyperparameters $\nu$ (left panel) and $\ell$ (right panel) represented by solid black curves, and the prior densities given by dotted red lines. Support of $\nu$ is $[0.5,1]$ and that for $\ell$ is $[0.1,1]$. Posterior samples were drawn using Metropolis--Hastings and utilizing $S$ obtained through Durbin's recursion.}}
	\label{post_hyp}
\end{figure}

Figure \ref{post_hyp} shows the posterior density plots of $\nu$ (left panel) and $\ell$ (right panel) and comparison with the uniform prior densities. Posterior samples were drawn using Metropolis--Hastings and the previously mentioned computation scheme. Posterior densities suggest that it is possible to learn the hyperparameters through this technique. Moreover, based on numerous simulation studies that we had conducted, the Metropolis--Hastings sampler for the hyperparameter update attained at least 15\% acceptance probability.

\section{Application to shape constrained estimation}\label{sec:app}

We now return to the constrained Gaussian regression setup in \eqref{eq:model}, and consider applications of our sampling algorithm to situations when $f$ is a smooth monotone or convex function. We first introduce some notation to define the basis functions employed by \cite{maatouk2017gaussian}. 

Let $\{u_j \in [0,1], \, j =0,1,\dots,N\}$ denote equally spaced knots on $[0, 1]$ with spacing $\delta_N = 1/N$ and $u_j = j/N$. Let
\begin{align*}
h_j(x) = h\bigg(\frac{x-u_j}{\delta_N}\bigg), \quad \psi_j (x) = \int_0^x h_j(t)\,dt, \quad\phi_j (x) = \int_0^x \int_0^t h_j(u)\,du\,dt; \quad x \in [0, 1]
\end{align*} 
where $h(x) = (1-|x|)\, \mathbbm{1}_{[-1, 1]}(x)$. The collection of functions $\{h_j\}$ is called the {\em interpolation basis} by \cite{maatouk2017gaussian}, since for any continuous function $f: [0, 1] \to \mb R$, the function $\widetilde{f}(\cdot)
= \sum_{j=0}^N f(u_j) \, h_j(\cdot)$ approximates $f$ by linearly interpolating between the function values at the knots $\{u_j\}$. 

The integrated basis $\{\psi_j\}$ and $\{\phi_j\}$ take advantage of higher-order smoothness. For example, if $f$ is continuously differentiable, then by the fundamental theorem of calculus, 
\begin{align*}
f(x) - f(0) = \int_0^x f'(t) dt. 
\end{align*}
Expanding $f'$ in the interpolation basis implies the model
\begin{align}\label{eq:m}
f(x) = \xi_0 + \sum_{j=0}^N \xi_{j+1} \psi_j(x). 
\end{align}
Similarly, if $f$ is twice continuously differentiable, we have 
$$
f(x) - f(0) -  x f'(0) = \int_0^x \int_0^t f''(s) \, ds dt. 
$$
Now expanding $f'$ and $f''$ in the interpolation basis implies the model 
\begin{align}\label{eq:c}
f(x) = \xi_0 + \xi^\ast x + \sum_{j=0}^N \xi_{j+1} \phi_j(x).
\end{align}
\cite{maatouk2017gaussian} showed that under \eqref{eq:m}, $f$ is monotone non-decreasing {\em if and only if} $\xi_i \ge 0$ for all $i = 1, \ldots, N+1$. Similarly, under \eqref{eq:c}, $f$ is convex non-decreasing {\em if and only if} $\xi_i \ge 0$ for all $i = 1, \ldots, N+1$. This equivalence relationship between the functional constraint and the linear inequality constraints on the basis coefficients is an attractive feature of the interpolation basis and isn't shared by many commonly used basis functions. 

For an unrestricted $f$, a GP prior on $f$ implies a dependent Gaussian prior on the coefficient vector $\xi = (\xi_1, \ldots, \xi_{N+1})^{\T}$.  A natural idea is to restrict this dependent prior subject to the linear restrictions on the coefficients which results in a dependent tMVN prior. Fitting the resulting model using a Gibbs sampler, the full conditional of $\xi$ assumes the form \eqref{full-cond-xi}, rendering our Algorithm~\ref{algo0} applicable. 

We provide more details regarding the model and prior for the monotone and convex cases separately. 
Let $X = (x_1, \ldots, x_n)^\T$ be the vector of $n$ design points, $Y = (y_1, \ldots, y_n)^\T$ be the vector of corresponding responses.


\subsection{Monotonicity constraint}\label{mon}
We can express \eqref{eq:m} in vector notation as
\begin{align}\label{model-mon}
Y = \xi_0 \mr 1_n + \Psi \xi + \varepsilon, \quad \varepsilon \sim \mc N_n(0, \sigma^2 \mr I_n), \quad \xi \in \mc C_{\xi}^{N+1}, 
\end{align}
where recall from \eqref{eq:c_xi} that $\mc C_{\xi}^{m}$ denotes the positive orthant in $\mb R^m$ and $\xi = (\xi_1, \ldots, \xi_{N+1})^{\T}$. Also, $\Psi$ is an $n \times (N+1)$ basis matrix with $i$th row $\Psi_i^\T$ where $\Psi_i = (\psi_0(x_i), \ldots, \psi_N(x_i))^\T$ and $\mr 1_n$ denotes an $n$ dimensional vector of all $1$'s.

%
The parameter $\xi_0 \in \mb R$ is unrestricted, and we place a flat prior $\pi(\xi_0) \propto 1$ on $\xi_0$. We place a tMVN prior on $\xi$ independently of $\xi_0$ as $p(\xi) \propto \mc N(\xi; 0, \tau^2 \, K) \, \ind_{\mc C_{\xi}}(\xi)$, where $K = (K_{jj'})$ with $K_{jj'} = k(u_j - u_j')$ and $k(\cdot)$ the stationary M{\'a}tern kernel with smoothness parameter $\nu > 0$ and length-scale parameter $\ell > 0$. To complete the prior specification, we place improper priors $\pi (\sigma^2) \propto 1/\sigma^2 \, ; \,\, \pi (\tau^2) \propto 1/\tau^2$ on $\sigma^2$ and $\tau^2$, and compactly supported priors $\nu \sim \mc U(0.5,1)$ and $\ell \sim \mc U(0.1,1)$ on $\nu$ and $\ell$. A straightforward Gibbs sampler is used to sample from the joint posterior of $(\xi_0, \xi, \sigma^2, \tau^2, \nu, \ell)$ whose details are deferred to the Appendix. The parameters $\sigma^2$, $\tau^2$ and $\xi_0$ have standard conditionally conjugate updates. The key feature of our algorithm is sampling the high-dimensional parameter $\xi$ using Algorithm~\ref{algo0} and updating $\nu$ and $\ell$ via Metropolis-within-Gibbs using Durbin's recursion as outlined in \S 2.3. 

Before concluding this section, we comment on a minor difference in our prior specification from \cite{maatouk2017gaussian}, which nevertheless has important computational implications. Since the basis coefficients $\xi_j, j \ge 1$ target the derivatives $f'(u_j)$, \cite{maatouk2017gaussian} consider a joint prior on $(\xi_0, \xi)$ obtained by computing the induced prior on $\big(f(0), f'(u_0), \ldots, f'(u_N) \big)$ from a GP prior on $f$, and then imposing the non-negativity restrictions. Since the derivative of a sufficiently smooth GP is again a GP, the joint distribution of $\big(f(0), f'(u_0), \ldots, f'(u_N) \big)$ can be analytically calculated. However, one downside is that the derivative of a stationary GP is no longer stationary in general, and thus sampling from the joint Gaussian prior of $\big(f(0), f'(u_0), \ldots, f'(u_N) \big)$ cannot take advantage of the embedding techniques for a stationary GP. We instead directly place a prior on $\xi$ induced from a stationary GP prior on $f'$ and then imposing the necessary restrictions. Since $\xi_0$ is only a single real-valued parameter, we break the dependence between $\xi_0$ and $\xi$ in the prior and assign a flat prior on $\xi_0$ independent of $\xi$. Our simulations suggest against any loss of efficiency in doing so, while there is a substantial computational gain because Algorithm~\ref{algo0} becomes readily applicable to update $\xi$ with our prior. 



\subsection{Convexity constraint}\label{con}
The development here proceeds in a similar fashion to the monotone case and we only provide a brief sketch. We can write \eqref{eq:c} in vector notation as 
\begin{align}\label{model-con}
 Y = \xi_0 \mr 1_n + \xi^\ast X + \Phi \xi + \varepsilon, \quad \varepsilon \sim \mc N_n(0, \sigma^2 \mr I_n), \quad \xi \in \mc C_{\xi}^{N+1}, 
\end{align}
where $\Phi$ is an $n \times (N+1)$ basis matrix with $i$th row $\Phi_i^\T$ and $\Phi_i = (\phi_0(x_i), \ldots, \phi_N(x_i))^\T$. The only additional parameter here from the previous case is $\xi^\ast$ to which we assign a flat prior on $\xi^\ast$ independent of everything else. For all other parameters, the exact same prior specification is employed as in the previous case. The details of the Gibbs sampler are once again deferred to the Appendix. Once again, the salient feature is the update of $\xi$ using Algorithm~\ref{algo0} and the usage of Durbin's recursion to update $\nu$ and $\ell$. 

%

\section{Simulation Results}\label{sec:sims}

\begin{figure}[htbp!]
\begin{center}
	\includegraphics[width=70mm]{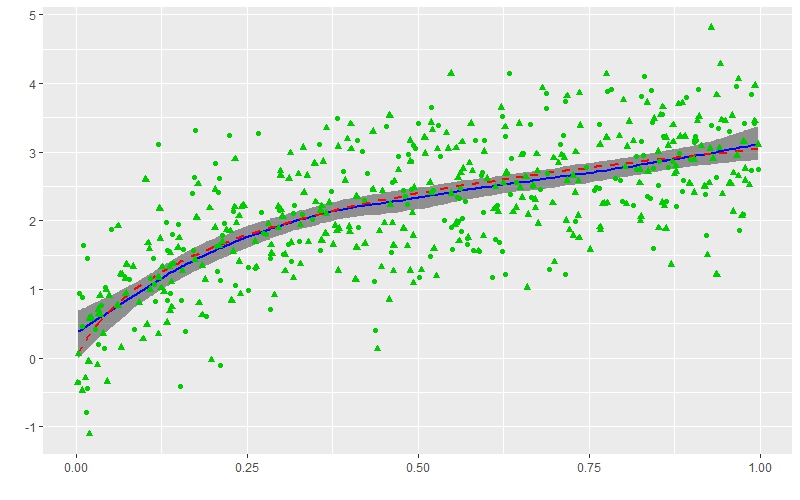}	
	\includegraphics[width=70mm]{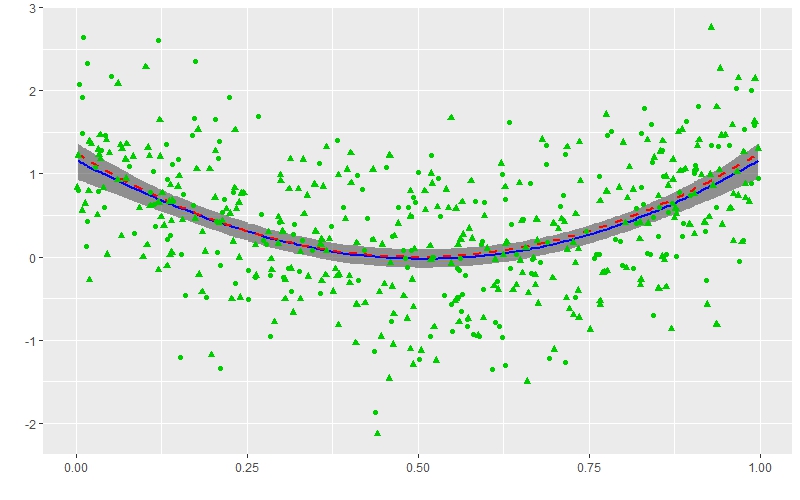}
		\caption{\small {\em Out of sample prediction at 200 held-out test data points for the monotone (left) and convex (right) examples. The green triangles and the green dots denote the training and held-out test data respectively. The dotted red line and the solid blue line respectively denote the true function and the point predictions at the test points, while the gray region provides a pointwise $95 \%$ credible interval.} \normalsize}
	 \label{fig:simul_illus}
	 \end{center}
	 \end{figure}

We now conduct a simulation study to provide empirical evidence that the proposed algorithm provides substantial speed-ups without sacrificing accuracy. We consider two examples corresponding to a monotone and convex truth respectively. For the monotone case, the true function $f(x) = \log(20x + 1)$, also considered in \cite{maatouk2017gaussian}, while in the convex case, the true $f(x) = 5 (x - 0.5)^2$. In both cases, we uniformly generated the covariates on $[0, 1]$ and added Gaussian noise. 
For all our simulations, we set $\eta = 50$ and the number of knots to be half the sample-size, $N = \lceil n/2 \rceil $. With $N = \lceil n/2 \rceil $, the computational complexity of Algorithm~\ref{algo0} within a single iteration of MCMC sampler is $\mc O \big( n^2 \big)$.
For the hyperparameters $\nu$ and $\ell$ inside the covariance kernel, we considered either of two situations where these were fixed {\em a priori} or updated inside the MCMC algorithm. When fixed, we used default choices of $\nu = 0.75$ and $\ell$ chosen so that the correlation at a maximum possible separation between the covariates equals $0.05$. When updated, we employed independent priors $\nu \sim \mc U(0.5,1)$ and $\ell \sim \mc U(0.1,1)$, as described earlier. Figure \ref{fig:simul_illus} provides an illustration of the point estimation along with uncertainty characterization using our Gibbs sampler for the monotone (left panel) and convex (right panel) cases. 

%

\subsection{Cost per-iteration}
We first illustrate the gains in per-iteration cost due to Algorithm~\ref{algo0} as opposed to directly sampling from a tMVN. We consider an alternative Gibbs sampler which samples $\xi \in \m C_{\xi}^{N+1}$ from its tMVN full-conditional using the Hamiltonian Monte Carlo (HMC) sampler of \cite{pakman2014exact}, implemented in the \textbf{{\fontfamily{qcr} \selectfont R}} package ``{\fontfamily{qcr}\selectfont tmg}". We kept $\nu$ and $\ell$ fixed and all other parameters are updated in the exact same way in either sampler. We did not consider the rejection sampler used by \cite{maatouk2017gaussian} as it becomes quite inefficient with increasing dimension, with the ``{\fontfamily{qcr}\selectfont tmg}" sampler substantially more efficient than the rejection sampler in high dimensions. The combination of  \cite{maatouk2017gaussian} with the 
``{\fontfamily{qcr}\selectfont tmg}" sampler does not exist in the literature to best of our knowledge, and thus we are being entirely fair in constructing the best possible competing method. 

\begin{figure}[htbp!]
\begin{center}
	\includegraphics[width=70mm]{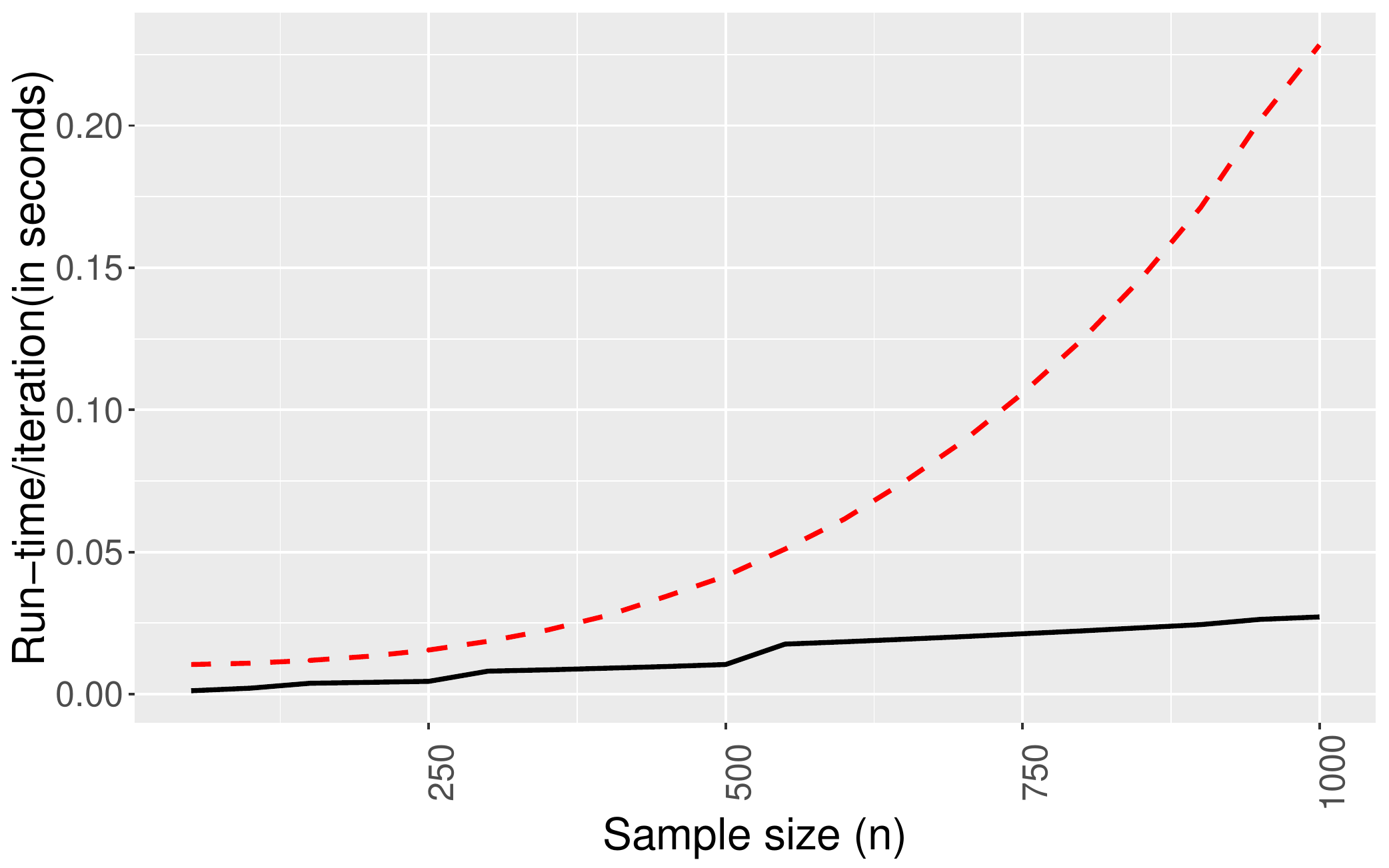}	\includegraphics[width=70mm]{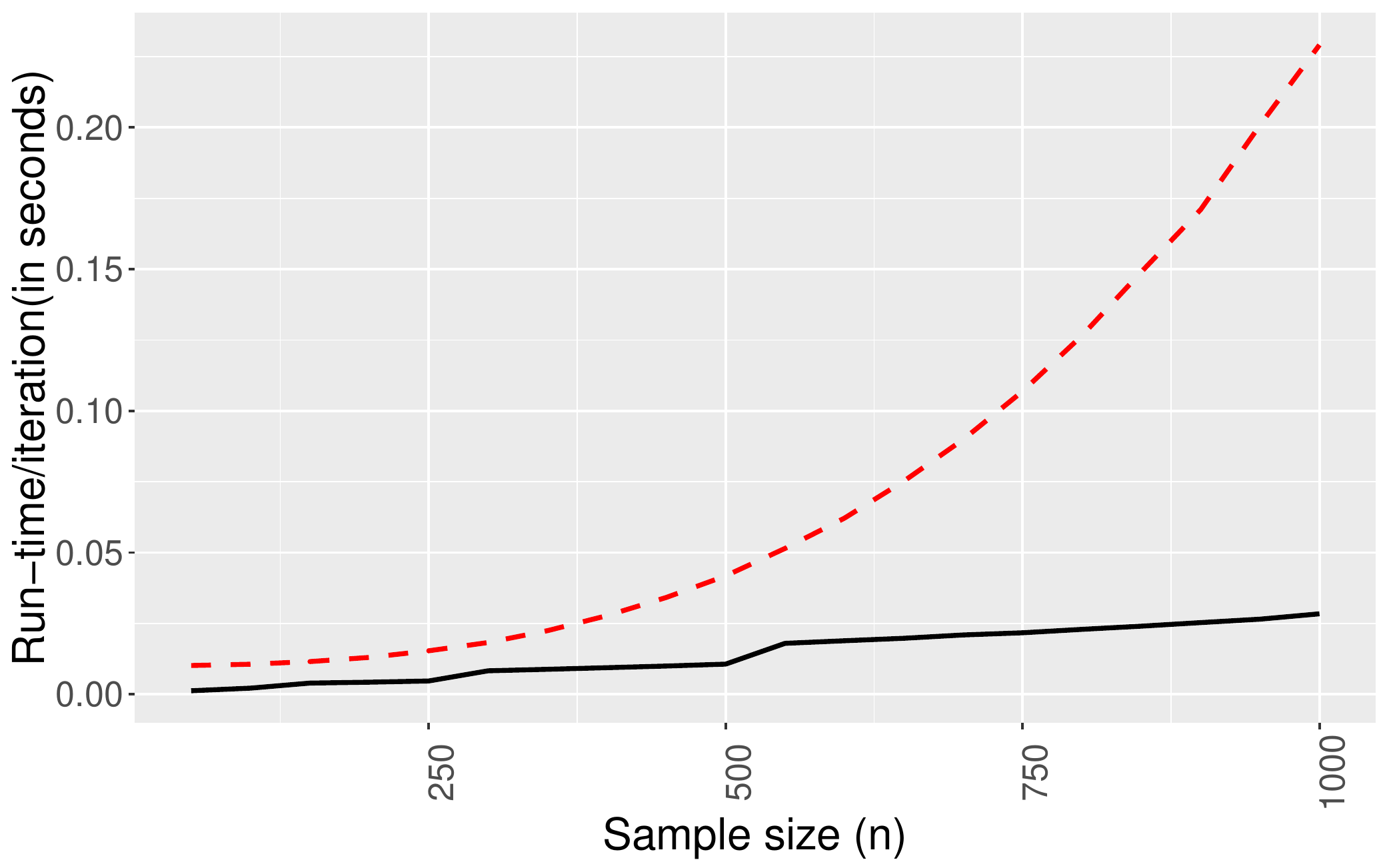}
		\caption{\small {\em Run-time per iteration (in seconds) against the sample size for two Gibbs samplers which only differ in the update of $\xi$ in the monotone (left panel) and convex (right panel) estimation context. Our Algorithm~\ref{algo0} is in solid black while the {\fontfamily{qcr}\selectfont tmg} sampler is in dotted red.} \normalsize}
	 \label{time-comp-mon-con}
	 \end{center}
	 \end{figure}

Figure \ref{time-comp-mon-con} plots the run-time per iteration (in seconds) against the sample size $n$ (varied between $50$ to $1000$ ) for the two approaches, both implemented under identical conditions on a quadcore Intel Core i7-2600 computer with 16 GB RAM. Evidently, Algorithm ~\ref{algo0} provides more pronounced improvements for larger $N$. While not shown here, one obtains a similar picture when the hyperparameters $\nu$ and $\ell$ are updated. Naturally, both methods incur additional computation time due to the additional calculations to compute the Metropolis--Hastings ratio. The ``{\fontfamily{qcr}\selectfont tmg}" sampler incurs additional burden in this case due to the continuous nature of the priors on $\nu$ and $\ell$, which either require storage of matrix inverses on a fine grid, or new matrix multiplications on the fly at each iteration. For fair comparison, while implementing ``{\fontfamily{qcr}\selectfont tmg}" along with hyperparameter updates, 
matrix inversions within each MCMC iterations were carried out using matrix inversion function of ``{\fontfamily{qcr}\selectfont FastGP}" that utilizes Trench's algorithm and requires $\mc O(N^2)$.

\subsection{Effective sample sizes}

We now investigate the mixing behavior of our Gibbs sampler by computing the effective sample size of the estimated function value at $75$ different test points. The effective sample size is a measure of the amount of the autocorrelation in a Markov chain, and essentially amounts to the number of independent samples in the MCMC path. From an algorithmic robustness perspective, it is desirable that the effective sample sizes remain stable across increasing sample size and/or dimension, and this is the aspect we wish to investigate here. We only report results for the monotonicity constraint; similar behavior is seen for the convexity constraint as well. 

\begin{figure}[htbp!]
\begin{center}
	\includegraphics[width=70mm]{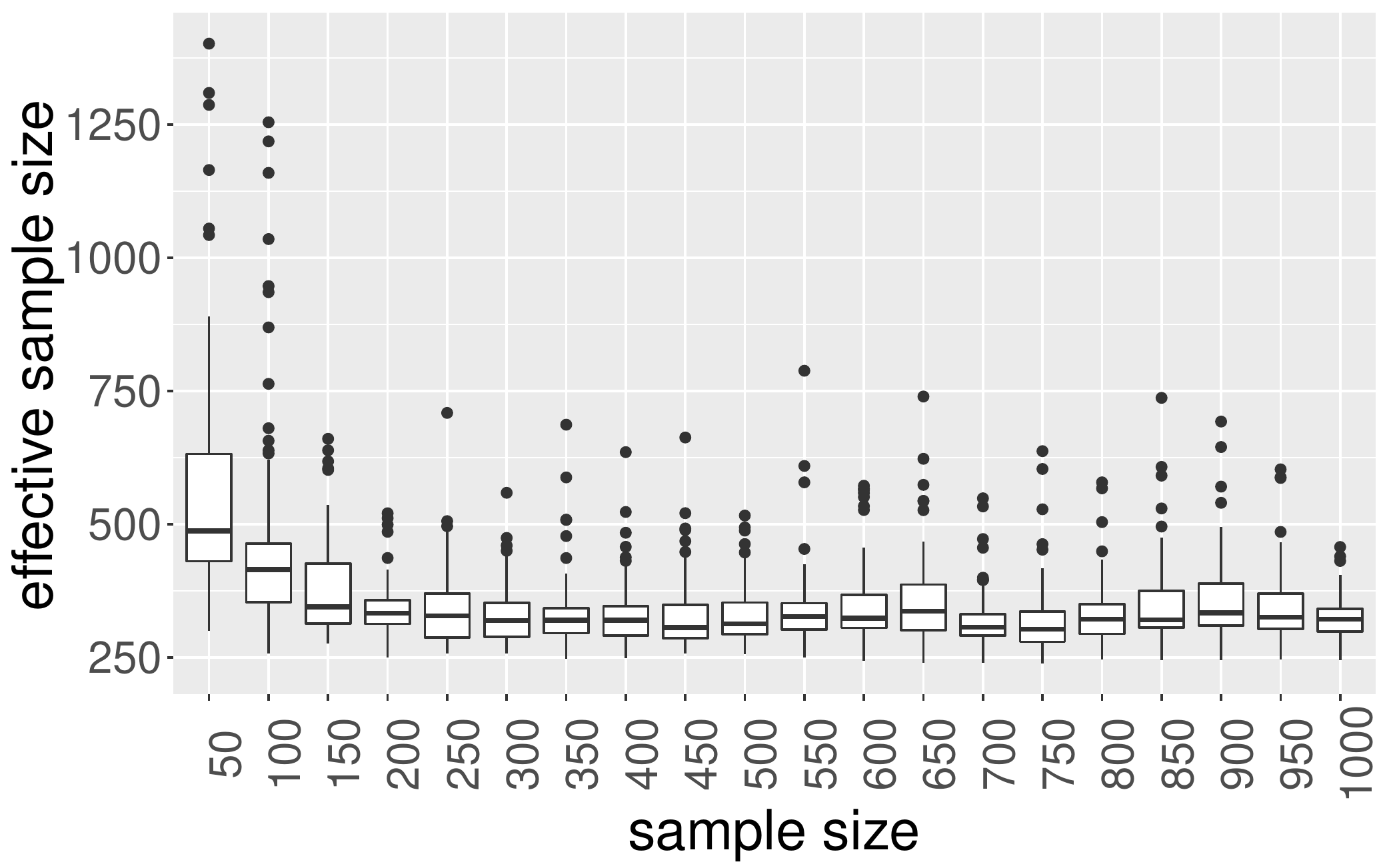}
	\includegraphics[width=70mm]{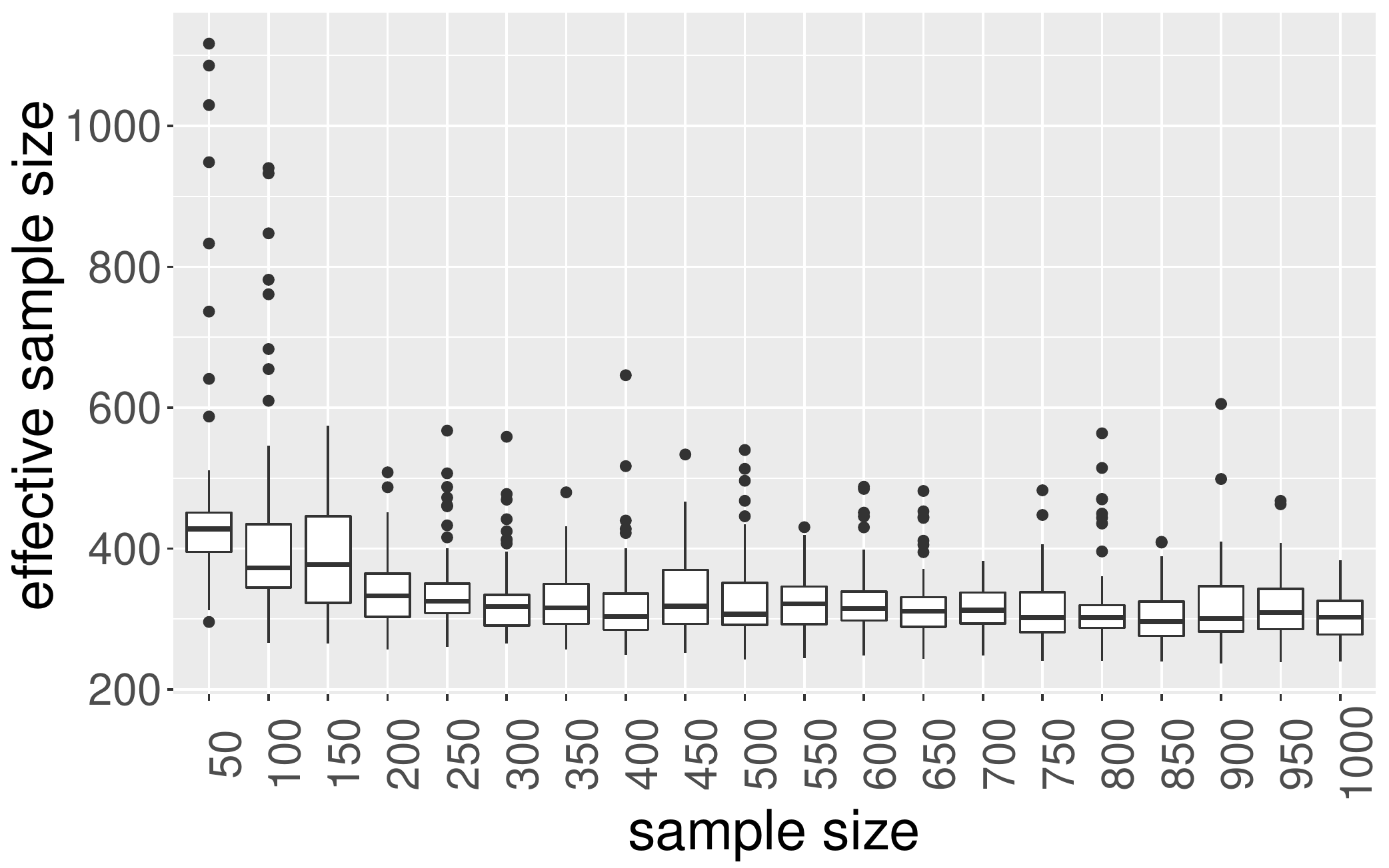}
	\caption{\small {\em  Boxplots of effective sample sizes of the estimated function value at $75$ different points for the monotone function estimation example. The effective sample sizes are calculated based on 10,000 MCMC runs and averaged over 5 random starting points. The left and right panels correspond to $\nu$ and $\ell$ fixed and updated within the MCMC, respectively.}\normalsize} \label{eff-nohyp}
\end{center}
\end{figure}

We consider 20 different values for the  sample size $n$ with equal spacing between $50$ and $1000$. Note that the dimension of $\xi$ itself grows between $25$ and $500$ as a result. For each value of $n$, we run the Gibbs sampler for 12,000 iterations with $5$ randomly chosen initializations. For each starting point, we record the effective sample size at each of the $75$ test points after discarding the first 2,000 iterations as burn-in, and average them over the different initializations. Figure \ref{eff-nohyp} shows boxplots of these averaged effective sample sizes across $n$; the left and right panel corresponds to $\nu$ and $\ell$ being fixed or updated within the MCMC, respectively. In either case, the effective sample sizes remain quite stable across growing $n$. When $\nu$ and $\ell$ are updated, there is a small dip in the median effective sample size which is pretty much expected. 


\section{Application on a real data set}\label{sec:real}

We considered a car price data\footnote{{\tiny available at \href{URL}{https://bitbucket.org/remcc/mbart/src/30f684da12ce5ad2d28a9b3c0a5fb166a5e15697/data/?at=master}}} studied by \cite{chipman2016high} on the sales prices of used Mercedes cars. The data consists of $1000$ observations on the car prices and we took mileage to be the explanatory variable. The relation between price and mileage is monotone decreasing, since a car with higher mileage is expected to have a lower price. 

\begin{figure}[H]
	\centering
	\begin{subfigure}{.5\textwidth}
		\centering
		\includegraphics[width=\textwidth]{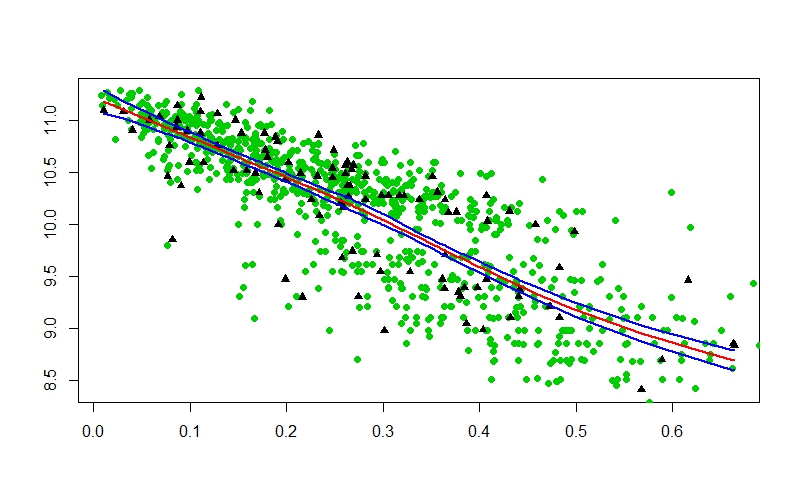}
	\end{subfigure}%
	\begin{subfigure}{.5\textwidth}
		\centering
		\includegraphics[width=\textwidth]{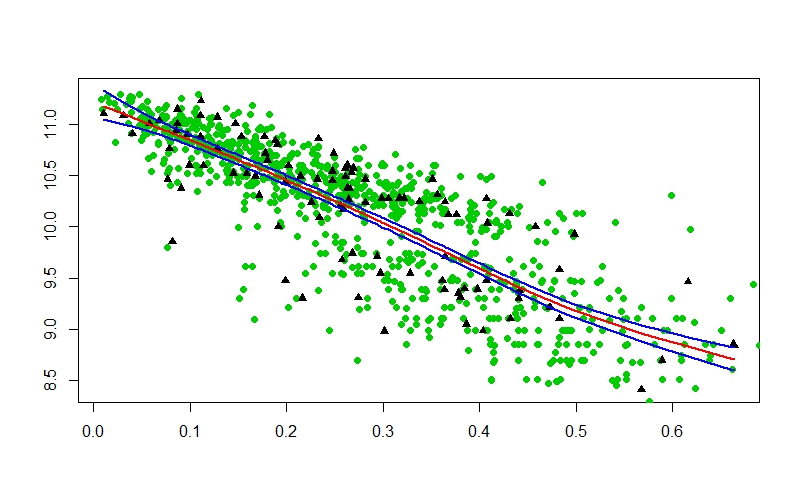}
	\end{subfigure}
	\caption{{\em Prediction accuracy for the cars price data using our method (left panel) and exact sampler (right panel) corresponding to the last fold of the validation. Red solid curve corresponds to the mean estimate, the region within two solid blue curves represent a 95\% credible interval, the green dots are training data points and the black triangles are test data points.}}
	\label{fig-real}
\end{figure}

We scaled the mileage to lie within $[0,1]$ and considered the log-transformed sales price as the response. As with the simulation study, we compared the performance of our algorithm to the exact algorithm which samples $\xi$ from its tMVN full conditional using the ``{\fontfamily{qcr}\selectfont tmg}" package. To compare the predictive performance, we considered 10 different random splits of the data into training and test tests. On each split, we reserved 900 training points for model fitting and the remaining 100 data points were held out for predictive evaluation. For each of the 10 datasets, both samplers were run for $6000$ iterations with the first $1000$ discarded as burn-in in each case. We also took into account the cases of fixing $\nu$ and $\ell$ or updating them within the MCMC. The average MSPE across the 10 splits for our method is $0.16645$ with standard error $0.0333$, while for the exact sampler the average MSPE is $0.16646$ with standard error $0.0332$ corresponding to the case where $\nu$ and $\ell$ were updated within the MCMC. As a visual illustration, Figure \ref{fig-real} shows the performance of both methods with hyperparameter updates, on the last fold of the $10$-fold validation process. While both algorithms resulted in similar accuracy, it took $0.25$ seconds to run one iteration of our algorithm in comparison to $0.39$ seconds for the exact sampler, corresponding to the case where $\nu$ and $\ell$ were updated in each iteration of the MCMC. For fixed $\nu$ and $\ell$, it took $0.014$ seconds to run one iteration of our algorithm in comparison to $0.055$ seconds for the exact sampler. Thus, as anticipated, our algorithm continues to multi-fold speed-ups without sacrificing on accuracy. 

\appendix
\section{Appendices}

\subsection{Full conditionals}

Consider model \eqref{model-mon} and the prior specified in section \ref{mon}. The joint distribution is given by:
$$ \pi (Y, \xi_0, \xi, \sigma^2, \tau^2) \propto \big(\sigma^2 \big)^{-\frac{n}{2} - 1} \exp \bigg\{ - \frac{1}{2 \sigma^2} \|Y - \xi_0 \mr 1_n - \Psi \xi \|^2 \bigg \} \\ \, \big(\tau^2 \big)^{-\frac{N+1}{2} - 1} \, \exp \bigg\{ - \frac{1}{2 \tau^2} \xi^\T K^{-1} \xi \bigg \} \, \ind_{\mc C_{\xi}}(\xi)  $$

Then,

$ \xi \mid Y, \xi_0, \sigma^2, \tau^2 $ is truncated multivariate Gaussian truncated on $\ind_{\mc C_{\xi}}(\xi)$.\\

$ \xi_0 \mid Y, \xi, \sigma^2, \tau^2 \sim \mc N(\bar{Y}^*, \sigma^2/n)$, where, $\bar{Y}^*$ is average of components of $Y^* = Y - \Psi \xi$. \\

$ \sigma^2 \mid Y, \xi_0, \xi, \tau^2 \sim \mc I \mc G \big( n/2, \|Y - \xi_0 \mr 1_n - \Psi \xi \|^2 /2 \big)$ \\

$ \tau^2 \mid Y, \xi_0, \xi, \sigma^2 \sim \mc I \mc G \big((N+1)/2, \xi^\T K^{-1} \xi /2 \big)$

\vspace{5mm}

Again, consider model \eqref{model-con} and the prior specified in section \ref{con}. The joint distribution is given by:
$$ \pi (Y, \xi_0, \xi_*, \xi, \sigma^2, \tau^2) \propto \big(\sigma^2 \big)^{-\frac{n}{2} - 1} \exp \bigg\{ - \frac{1}{2 \sigma^2} \|Y - \xi_0 \mr 1_n - \xi_* X - \Phi \xi \|^2 \bigg \} \big(\tau^2 \big)^{-\frac{N+1}{2} - 1} \exp \bigg\{ - \frac{1}{2 \tau^2} \xi^\T K^{-1} \xi \bigg \} \, \ind_{\mc C_{\xi}}(\xi)  $$

Then,

$ \xi \mid Y, \xi_0, \xi_*, \sigma^2, \tau^2 $ is truncated multivariate Gaussian truncated on $\ind_{\mc C_{\xi}}(\xi)$.\\

$ \xi_0 \mid Y, \xi_*, \xi, \sigma^2, \tau^2 \sim \mc N(\bar{Y}^*, \sigma^2/n)$, $\bar{Y}^*$ is average of components of $Y^* = Y - \xi_* X - \Phi \xi$. \\

$ \xi_* \mid Y, \xi_0, \xi, \sigma^2, \tau^2 \sim \mc N( \sum_{i=1}^{n} x_i y_i^{**}/ \sum_{i=1}^{n} x_i^2, \sigma^2/ \sum_{i=1}^{n} x_i^2)$, where $Y^{**} = Y - \xi_0 \mr 1_n - \Phi \xi$. \\

$ \sigma^2 \mid Y, \xi_0, \xi_*, \xi, \tau^2 \sim \mc I \mc G \big( n/2, \|Y - \xi_0 \mr 1_n - \xi_0 X - \Phi \xi \|^2 /2 \big)$ \\

$ \tau^2 \mid Y, \xi_0, \xi_*, \xi, \sigma^2 \sim \mc I \mc G \big( (N+1)/2, \xi^\T K^{-1} \xi /2 \big)$\\

Algorithm~\ref{algo0} was used to draw samples from the full conditional distribution of $\xi$ while sampling from the full conditionals of $\xi_0$, $\xi_*$, $\sigma^2$ and $\tau^2$ are routine. 

\subsection{R code}

We used \textbf{{\fontfamily{qcr} \selectfont R}} for the implementation of Algorithm~\ref{algo0} and Durbin's recursion to find the inverse of the Cholesky factor, with the computation of the inverse Cholesky factor optimized with \textbf{{\fontfamily{qcr} \selectfont Rcpp}}. We provide our code for implementing the monotone and convex function estimation procedures in \S \ref{sec:sims} and \S \ref{sec:real} in the Github page mentioned in \S \ref{sec:intro}. There are six different functions to perform the MCMC sampling for monotone increasing, monotone decreasing, and convex increasing functions with and without hyperparameter updates. Each of these main functions take $x$ and $y$ as inputs along with other available options, and return posterior samples on $\xi_0$, $\xi^*$, $\xi$, $\sigma$, $\tau$ and $f$ along with posterior mean and 95\% credible interval of $f$ on a user-specified grid. A detailed description on the available input and output options for each function can be found within the function files.


\bibliographystyle{plainnat}
\bibliography{refs_NPC}

\end{document}